# Noisy-threshold control of cell death

**Jose M. G. Vilar**[1,2§]

[1]Biophysics Unit (CSIC-UPV/EHU) and Department of Biochemistry and Molecular Biology, University of the Basque Country, P.O. Box 644, 48080 Bilbao, Spain

[2]IKERBASQUE, Basque Foundation for Science, 48011 Bilbao, Spain

[§]Corresponding author

Email address:

    JMGV: j.vilar@ikerbasque.org

# Abstract

**Background**

Cellular responses to death-promoting stimuli typically proceed through a differentiated multistage process, involving a lag phase, extensive death, and potential adaptation. Deregulation of this chain of events is at the root of many diseases. Improper adaptation is particularly important because it allows cell sub-populations to survive even in the continuous presence of death conditions, which results, among others, in the eventual failure of many targeted anticancer therapies.

**Results**

Here, I show that these typical responses arise naturally from the interplay of intracellular variability with a threshold-based control mechanism that detects cellular changes in addition to just the cellular state itself. Implementation of this mechanism in a quantitative model for T-cell apoptosis, a prototypical example of programmed cell death, captures with exceptional accuracy experimental observations for different expression levels of the oncogene Bcl-$x_L$ and directly links adaptation with noise in an ATP threshold below which cells die.

**Conclusions**

These results indicate that oncogenes like Bcl-$x_L$, besides regulating absolute death values, can have a novel role as active controllers of cell-cell variability and the extent of adaptation.

# Background

Cells in multicellular organisms have the ability to actively control their own death and engage in an organized self-destruction process known as programmed cell death, or apoptosis. This ability is necessary in many situations, ranging from embryonic development to the maintenance of mature organisms, and its deregulation is a hallmark of cancer [1] and many other diseases [2].

Apoptosis can be triggered by a variety of internal or external stimuli, such as specific death signals, lack of growth factors, DNA damage, and different types of stress [3]. These stimuli frequently converge with the metabolic state of the cell into the mitochondrial pathway, which promotes apoptosis through activation of the caspase cascade by releasing cytochrome c into the cytosol [3-5]. Low intracellular ATP levels favor the release of cytochrome c, resulting in high apoptotic rates [6, 7]. In this way, the absence of nutrients or signals that stimulate metabolism triggers apoptosis in a wide variety of cell types (Figure 1A).

The coupling of metabolism with apoptosis is remarkably important in cancer, especially because many anticancer drugs targeted against growth promoting pathways selectively induce apoptosis in cells with highly upregulated metabolism [8-10]. Extensive work has been done along these lines to decipher how different cellular components are wired to coordinate apoptotic responses [11, 12]. The connection of the single cell behavior with the collective dynamics of cell populations, however, has remained largely unexplored.

It is still not clear how the death-or-alive binary decision of single cells leads to the observed analog response of the cell population, where death increases continuously with decreasing ATP levels [13]. This connection is particularly challenging because classic population dynamics approaches [14] would require the death rate to peak at an intermediate ATP level to account for adaptation at low ATP levels, which paradoxically would imply that low ATP levels strongly favor survival. Here, the single-cell and population scales are connected through a quantitative mathematical model that identifies cellular variability, or noise, as the key element.

## Methods

In general, cells have different levels of ATP, of proteins that promote or prevent cytochrome c release, and of the components of the caspase cascade. The most straightforward assumption is to consider that for each individual cell there is a specific value of the ATP level, an ATP level threshold, below which apoptosis is triggered and that intercellular variability in the cellular components makes this threshold noisy. Thus, a cell will die whenever the threshold is crossed, either because its ATP level decreases below its threshold or, vice versa, because its threshold increases over its ATP level.

In this paper, the focus is on death promoting stimuli that strongly affect metabolism, such as growth factor or nutrient withdrawal. Under these conditions, intracellular ATP levels change much faster than thresholds do and it is assumed that the noise in the threshold is quenched. Explicitly, the threshold is considered to be a time-independent random variable with precise statistical properties (Figure 1B).

The connection between the threshold distribution and cell death can be expressed in mathematical terms through the survival ATP-function, $S(a)$, which is defined as the fraction of cells with a threshold below the average intracellular ATP level, denoted here by $a$. The initial threshold distribution is therefore given by $T(x) = \frac{dS(a)}{da}\Big|_{a=x}$, where $x$ is the ATP threshold.

The apoptotic contribution to the death rate is the rate at which cells cross the threshold. Taking into account that the total number of cells is proportional to the fraction of cells that survive, the death rate when the ATP level decreases with time is given by

$$r(a, \frac{da}{dt}) = -\frac{1}{S(a)} \frac{dS(a)}{da} \frac{da}{dt}, \qquad (1)$$

which uses the chain rule for derivatives: $dS(a)/dt = (dS(a)/da)(da/dt)$. When the ATP level increases with time, within this model, the threshold would never be crossed and the apoptotic death rate would be zero.

A far-reaching implication of the threshold crossing assumption is that the death rate is not a function of just the ATP level but also of the ATP rate of change: $da/dt$.
Thus, a precise testable prediction of the model is that changing the ATP level from decreasing ($da/dt < 0$) to increasing ($da/dt > 0$) will bring the apoptotic death rate to zero, irrespective of the value of the ATP level itself.

The effects of apoptosis in response to growth factor or nutrient withdrawal in a population of cells are modeled by considering the intracellular ATP level, $a$, and the number of cells, $N$, as dynamic variables together with the net consumption of ATP per cell, $f(a)$, and the net death rate, $r(a, da/dt)$:

$$\frac{da}{dt} = -f(a), \qquad (2)$$

$$\frac{dN}{dt} = -r(a, \frac{da}{dt})N. \tag{3}$$

This model implements that the population dynamics response to energy depletion depends on both the energetic state of the cell and its dynamics. The complexity of cell death regulation [12] is collapsed into the form of the function $r(a, da/dt)$, which can be inferred from experimental data, as shown below.

## Results

This approach can be applied straightforwardly to analyze experimental data for T-cell apoptosis upon growth factor withdrawal. Following loss of receptor engagement, T-cells rapidly downregulate the glucose transporter, glut1, which leads to lower intracellular ATP levels and results in apoptosis [15]. Experimental data is available for three levels of expression of the antiapoptotic protein Bcl-$x_L$ [15]. For the three levels of expression of Bcl-$x_L$, the average ATP levels follow very similar temporal dynamics, thus indicating that Bcl-$x_L$ does not affect cellular metabolism substantially.

The cellular metabolism experimental data (Figure 2A) can be incorporated into the model by considering that the ATP level, $a(t)$, decays exponentially from its initial value, $a(0)$, to a basal value, $ATP^*$, with a characteristic time $\tau_{ATP}$, which implies a net ATP consumption given by

$$f(a) = (a - ATP^*) / \tau_{ATP}. \tag{4}$$

In this case, autophagy provides the energy that allows cells to maintain a basal metabolic state in the absence of nutrient uptake [16].

The effects of Bcl-$x_L$ on apoptosis are taken into account by the shape of the threshold distribution. Under fairly general conditions, when multiplicative stochastic effects are present, such as stochastic reaction rates, the statistics usually follows a lognormal distribution [17, 18]. Therefore, a natural choice for the survival ATP-function is

$$S(a) = \int_0^a \frac{1}{\sqrt{2\pi\sigma}} \frac{1}{x} e^{-\frac{(\log(x)-\mu)^2}{2\sigma}} dx, \qquad (5)$$

where here $x$ is the threshold, and $\mu$ and $\sigma$ are two parameters related to the mean, $\bar{x} = e^{\mu+\sigma/2}$, and variance, $\overline{x^2} - \bar{x}^2 = (e^\sigma - 1)e^{2\mu+\sigma}$, of the distribution. The coefficient of variation is given by $(\overline{x^2} - \bar{x}^2)^{1/2} / \bar{x} = (e^\sigma - 1)^{1/2}$ and can be viewed as a measure of the cell-death noise. In general, $\mu$ and $\sigma$ depend on the experimental conditions, cell type, and expression levels of Bcl-$x_L$.

## Discussion

The model defined by Equations (1)–(5) accurately accounts for the observed apoptotic behavior in response to growth factor withdrawal (Figure 2B) when the two parameters that characterize the threshold distribution are appropriately chosen for each cell type.

The response includes a lag phase, extensive death, and adaptation. All these features, in contrast to the intracellular ATP dynamics, strongly depend on Bcl-$x_L$ expression. For the lowest expression, the cell type "Bcl-$x_L$" shows substantial death as soon as the ATP level starts to drop, with a very short lag phase (~ 5 hours). In this case, there is no adaptation and after 25 hours the whole cell population is dead. In the case of the cell type "Bcl-$x_L$ 1E1", the lag phase lasts 20 hours approximately and is succeeded by a period of substantial death (until a time around 75 hours) before adaptation settles in with a small subpopulation of surviving cells. With the highest expression, the cell type "Bcl-$x_L$ 4.1" is very resistant to energy depletion. It has virtually no cell death until 40 hours after growth factor withdrawal and a large fraction of cells are able to survive.

The initial distributions inferred from the experimental data (Figure 3), by choosing the parameters $\mu$ and $\sigma$ to reproduce the observed behavior, indicate that, as Bcl-$x_L$ expression increases, the thresholds shift on average toward lower values of the ATP level, consistent with the well-known antiapoptotic properties of Bcl-$x_L$. At the same time, the distributions increase their relative spread, resulting in higher cell-to-cell variability for higher Bcl-$x_L$ expression (Figure 4).

It is important to emphasize that other types of threshold distributions, besides lognormal, could also be used in the model. However, if they are able to accurately reproduce the experimental data, they will look very similar to the lognormal distributions of Figure 3 in the range where intracellular ATP changes (from 3.54 to 0.32 fmol/cell). The reason is that in this system the shape of the threshold distribution is completely determined by the death rate. This result follows straightforwardly from integration of Eq. (1), which leads to the relationship $\int_{x_1}^{x_2}(r/(da/dt))da = \ln(S(x_1)/S(x_2))$. For instance, a Gaussian distribution would be able to provide relatively accurate results for the cell type "Bcl-$x_L$" but would fail to do so for the other two cell types. A key feature of the lognormal distribution is its ability to recapitulate the experimental observations for the three cell types.

The threshold distribution of live cells evolves in time because cells die when their threshold is higher than the ATP level (Figure 5 and Additional File 1). For monotonously decreasing ATP levels, this time-dependent behavior can be expressed explicitly in mathematical terms as $T(x,t) = (dS(x)/dx)\Theta(a(t) - x)$, where $\Theta(a(t) - x)$ is the Heaviside step function, which is 1 when $a(t) - x > 0$ and 0, otherwise. The resulting distribution is normalized to the fraction of surviving cells, $\int_0^\infty T(x,t)dx = S(a(t))$.

The results of the model indicate that the increased variability of cells that overexpress Bcl-$x_L$ leads to highly heterogeneous responses, allowing a subpopulation of cells to escape death. Thus, within this model, adaptation observed in the late stages of the response to energy depletion is not the result of the apoptotic machinery adapting to low ATP levels but the result of the combined effects of high threshold variability with intracellular ATP reaching a steady state. If there were no noise, the whole population would either die or survive. When all the cells die, there is no adaptation. When all the cells survive, there is no response to the death stimulus and therefore there is no adaptation either. Thus, control of the population dynamics requires controlling not only the average values of the factors that control death but also their statistical properties.

The precise details of the transition from the single-cell to the cell-population dynamics can be elucidated by considering finite populations explicitly (Figure 6). Mathematically, the number of live cells is given by $N_{live}(t) = \sum_{i=1}^{N} \Theta(a(t) - x_i)$, where $x_i$ is the threshold of the cell $i$ of a population of $N$ initial (live) cells. The threshold $x_i$ is a random variable with the statistical properties given by the distribution for each cell type. At the single cell level, for $N=1$, there is no adaptation: either the cell dies at some random time or survives. For small populations, whether there is adaptation depends on the realization. There is no adaptation if all or none of the cells survive. In the limit of large $N$, the continuous population dynamics, $N_{live}(t)/N \approx S(a(t))$, is recovered.

## Conclusions

Recently, there has been an increasing interest on the role that molecular noise can play on the cellular behavior. Noise, in the form of random fluctuations in the number of molecules has been observed in many cellular processes, such as gene expression [19-22], protein abundance regulation [23-25], intracellular oscillations [26-28], and behavioral variability [29, 30]. The results presented here indicate that the oncogene Bcl-$x_L$ exploits the inherent stochastic nature of molecular events to generate cell-death variability, which transforms a single-cell discrete event like death into a graded population response and leads to adaptation.

These two noise-induced features have far-reaching physiological consequences. On the one hand, a graded death response is essential for a smooth control of total cell numbers in multicellular organisms [13, 31, 32]. On the other hand, adaptation is a key mechanism that allows a subpopulation of cells to survive in the face of death-promoting stimuli, leading to the eventual failure of many anticancer therapies [33]. In the case of T-cell apoptosis, these two apparently disconnected features are in fact two sides of a single control mechanism based on noisy thresholds to trigger death.

## Authors' contributions

JMGV performed the research and wrote the manuscript. All authors read and approved the final manuscript.

## Acknowledgements

This work was supported by the MICINN under grant FIS2009-10352.

# Figures

**Figure 1 - Noise-threshold model of apoptosis**

(A) The cartoon illustrates the cellular processes that serve as a basis for the model. The metabolic state of the cell determines the intracellular ATP level. The ATP level controls negatively the release of cytochrome c (cyt c) from mitochondria into the cytosol, which promotes apoptosis. Antiapoptotic proteins such as Bcl-$x_L$ prevent the release of cytochrome c and can compensate for low ATP levels. (B) The model assumes that each individual cell has a given threshold for the ATP level that separates death from survival. Cells that survive, emphasized in blue, are those with thresholds below the ATP level (set here at ~1 fmol/cell for illustrative purposes).

**Figure 2 - Model results for three different levels of expression of the oncogene Bcl-$x_L$**

(A) ATP level as a function of time after IL-3 growth factor removal. (B) Fraction of surviving cells for different time lags after IL-3 removal. Different symbol shapes and colors refer to the same cell types in both panels A and B. They correspond to experimental data from Ref. [15] for three cell types, labeled as "Bcl-$x_L$", "Bcl-$x_L$ 1E1", and "Bcl-$x_L$ 4.1", which have low, high, and higher expression of Bcl-$x_L$, respectively. Continuous lines correspond to model results from Equations (1)–(5) with $a(0) = 3.54$ fmol/cell, $ATP^* = 0.32$ fmol/cell, and $\tau_{ATP} = 19.9$ hours for all three cell types. The inferred values of the pairs $(\mu, \sigma)$ for the cell types "Bcl-$x_L$", "Bcl-$x_L$ 1E1", and "Bcl-$x_L$ 4.1" are (0.63, 0.074), (-0.38, 0.27), and (-1.53, 0.58), respectively.

**Figure 3 - Threshold distributions**

The inferred ATP threshold distributions are shown for three different levels of expression of the oncogene Bcl-$x_L$. The values of the parameters $\mu$ and $\sigma$ that characterize the threshold distributions, given by $T(x) = \frac{1}{\sqrt{2\pi}\sigma} \frac{1}{x} e^{-\frac{(\log(x)-\mu)^2}{2\sigma}}$, are the same as in Figure 2.

**Figure 4 - Cell-death noise**

The cell-death noise, defined as the standard deviation over the mean of the ATP threshold distribution, increases with the expression of Bcl-$x_L$. Its inferred value for the cell types "Bcl-$x_L$", "Bcl-$x_L$ 1E1", and "Bcl-$x_L$ 4.1" is 0.28, 0.56, and 0.88, respectively. The mean (fmol/cell), variance (fmol$^2$/cell$^2$), and noise (dimensionless) are shown as a function of Bcl-$x_L$ expression normalized by that of the cell type "Bcl-$x_L$". Bcl-$x_L$ expression reported in Western blots of Ref. [15] was quantified with the program ImageJ (http://rsbweb.nih.gov/ij/).

**Figure 5 - Time evolution of the threshold distribution and adaptation**

The ATP threshold distribution for the cell type "Bcl-$x_L$ 1E1" is shown at different time points after IL-3 growth removal together with the dynamics of the intracellular ATP level (red thick line). The distribution is normalized to the survival fraction $S(a(t))$. The values of the parameters of the model are the same as in Figure 2. In this case, adaptation observed in the late stages of the response is the result of the combined effects of high threshold variability with intracellular ATP adapting to a new steady state.

**Figure 6 - From single-cell to collective behavior**
Representative survival curves are shown for populations of 1, 10, 100, and 1000 cells. The number of live cells is computed from $N_{live}(t) = \sum_{i=1}^{N} \Theta(a(t) - x_i)$, where $x_i$ is the threshold of the cell $i$ of a population of $N$ initial (live) cells. The threshold $x_i$ is chosen randomly from the distribution for the cell type "Bcl-$x_L$ 1E1" inferred from the experimental data (Figure 2).

## Additional files
**Additional file 1 – Threshold distribution dynamics.**
The Flash Movie shows an animation of the threshold distribution dynamics for the cell type "Bcl-xL 1E1" as in Figure 5.

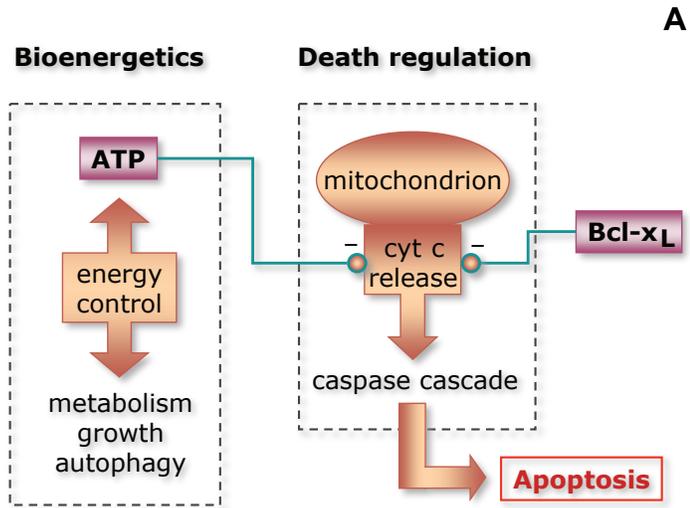
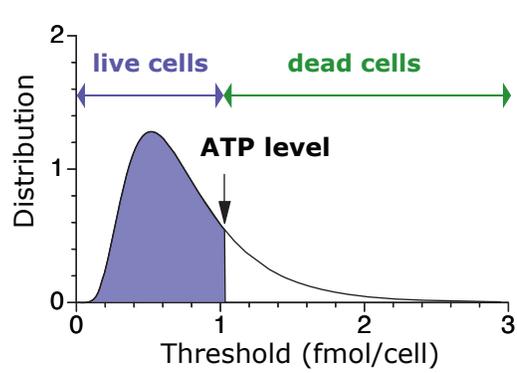

**Figure 1**

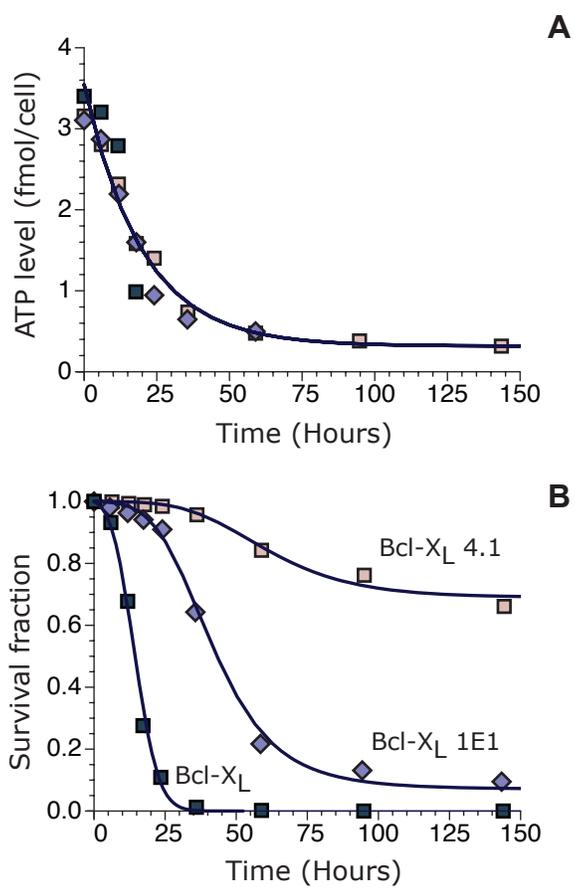

**Figure 2**

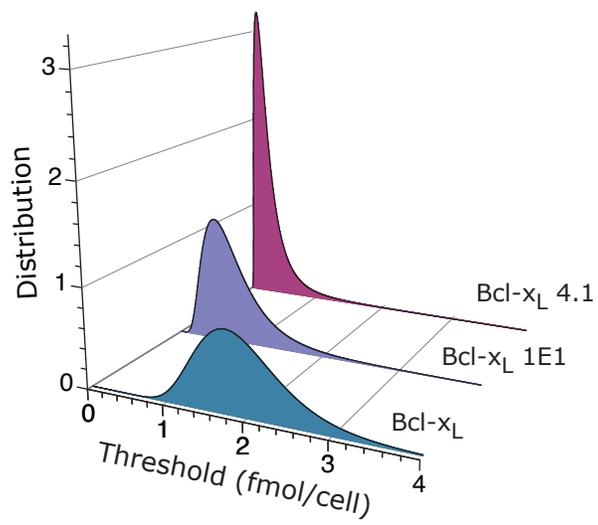

**Figure 3**

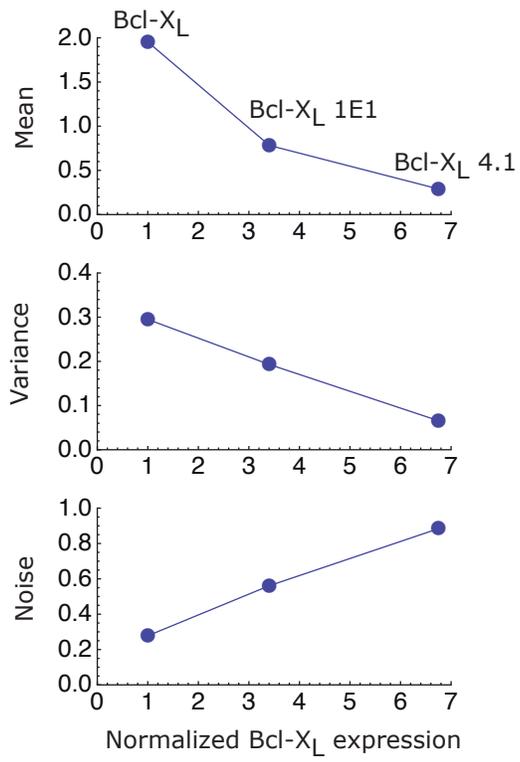

**Figure 4**

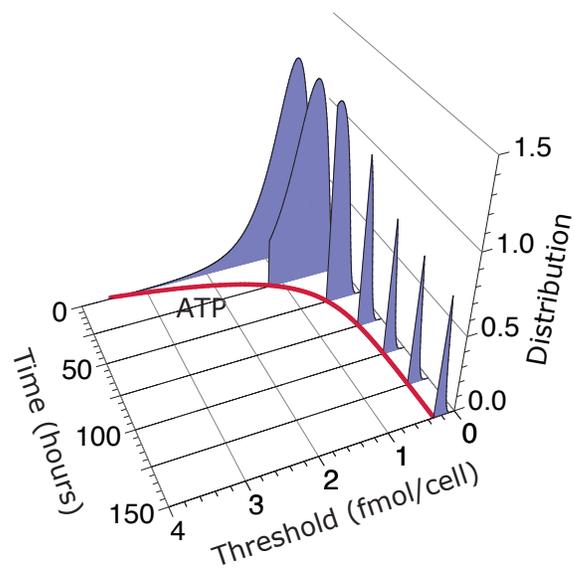

**Figure 5**

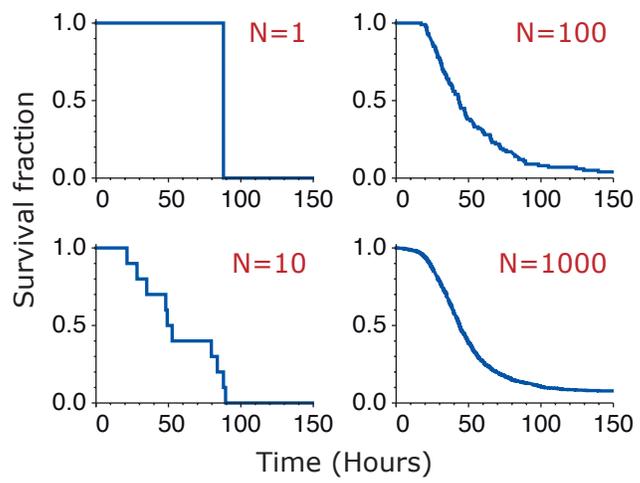

**Figure 6**